\documentclass{appolb}
\usepackage{graphicx}

\def\be{\begin{equation}}
\def\ee{\end{equation}}
\def\bea{\begin{eqnarray}}
\def\eea{\end{eqnarray}}

 \usepackage{epsfig}
 \usepackage{epstopdf}

\begin{document}
\title{Decay of the Higgs boson $h\to\tau^- \tau^+\to\pi^-\nu_{\tau}\,\pi^+{\bar \nu}_\tau$
for a non-Hermitian Yukawa interaction}

\author{Alexander~Yu.~Korchin
\address{NSC Kharkiv Institute of Physics and Technology, 61108 Kharkiv, Ukraine } 
\address{V.N.~Karazin Kharkiv National University, 61022 Kharkiv, Ukraine } 
\\
\vspace{0.25cm}
Vladimir~A.~Kovalchuk\footnote{Deceased} 
\address{NSC Kharkiv Institute of Physics and Technology, 61108 Kharkiv, Ukraine}
}

\maketitle
\begin{abstract}
The differential rate of the decay of the Higgs boson ($h$) to a pair of $\tau $ leptons with their subsequent decay in the 
$\tau^-\to \pi^- \nu_\tau$ and $\tau^+\to \pi^+{\bar \nu}_\tau$ channels is studied.
The Yukawa interaction between the Higgs boson and the $\tau$ leptons is assumed to include scalar ($S$) and pseudoscalar ($PS$) couplings.
Angular distributions of the pions in the  $h\to\tau^- \tau^+\to\pi^-\nu_{\tau}\,\pi^+{\bar \nu}_\tau$  decay are considered.
For real values of the $S$ and $PS$ couplings, this decay is known to be a source of information on $\mathcal{CP}$ violation in
the $h \tau \tau $ interaction.  In the present paper, the main attention is paid to a possible non-Hermiticity of this interaction.
Influence of non-Hermiticity on the distribution of the angle between planes of the $ \tau^- \to \pi^-\nu_{\tau}$
and $\tau^+ \to \pi^+ {\bar \nu}_{\tau}$ decays, and distribution of the polar angle of one of the pions are analyzed.  Asymmetries
sensitive to parameters of $\mathcal{CP}$ violation and non-Hermiticity of $h \tau \tau$ interaction  are proposed.
 \end{abstract}
%


\section{Introduction}
\label{sec:Introduction}

In framework of the Standard Model (SM), the fermion masses are generated
by the Yukawa interaction between the Higgs field and fermion fields.
Measurement of the corresponding couplings is needed for identification of the
particle $h$ with the Higgs boson.  In the SM, the Higgs boson is the $\mathcal{CP}$-even scalar particle.
However, there exist many models with a more complicated structure of the Higgs sector in which
both the $\mathcal{CP}$-even and $\mathcal{CP}$-odd scalar particles can exist, as well as
particles which do no have definite $\mathcal{CP}$ parity (see recent review 
\cite{LHCHiggsCrossSectionWorkingGroup:2016ypw} and references
therein). Therefore, it is possible that the observed Higgs boson $h$ \cite{Aad:2012tfa,Chatrchyan:2012ufa}
is a mixture of  $\mathcal{CP}$-even and $\mathcal{CP}$-odd states, although
other possibilities are not excluded.
Thus, clarification of the $\mathcal{CP}$ properties of the Higgs boson is a necessary step in investigation 
of the mechanism which breaks electroweak symmetry and generates the particle masses.
The present status of the LHC measurements of the $\mathcal{CP}$ properties of the Higgs-boson interactions 
with vector bosons and fermions is reviewed in Ref.~\cite{Bass:2021acr}.

In the SM, the source of violation of the $\mathcal{CP}$ symmetry is unremovable phase 
in the Cabibbo-Kobayashi-Maskawa (CKM) mixing matrix~\cite{Cabibbo:1963yz,Kobayashi:1973fv}. Moreover,
the existing data indicate that this phase is the dominant source of $\mathcal{CP}$ violation in the flavor changing processes. However,
model calculations show that $\mathcal{CP}$ violation in the SM is too small to explain the matter-antimatter asymmetry in the
Universe~\cite{Farrar:1993hn,Davidson:2008bu}. There should be other sources of $\mathcal{CP}$ violation beyond the CKM mechanism. 
Thus, the search for new sources of $\mathcal{CP}$ violation is one of the main directions in the particle physics. 
One of possibilities in this search is the Higgs boson decay $h\to\tau^-\tau^+$.

The study of the $\mathcal{CP}$ properties and violation of the $\mathcal{CP}$ symmetry in the Higgs sector, 
using the correlations between the spins of $\tau$ leptons created in the Higgs-boson decay, has been carried 
out in a series of papers, {\it e.g.}
\cite{DellAquila:1988bko,Kramer:1993jn,Grzadkowski:1995rx,Bernreuther:1997af,
Bower:2002zx,Desch:2003mw,Worek:2003zp,Desch:2003rw,Rouge:2005iy,Berge:2008wi,Berge:2008dr,Berge:2011ij,
Harnik:2013aja,Berge:2013jra,Berge:2014sra,Askew:2015mda,Berge:2015nua,Korchin:2016rsf,Chen:2017bff,Hagiwara:2016zqz,
Jeans:2018anq,Chen:2017nxp,Ge:2020mcl}.

Another important aspect of the Yukawa interaction is Hermiticity of the Lagrangian.
In the SM, the Lagrangian of the interaction between fermions and scalar fields satisfies the 
symmetry with respect to the gauge transformations $SU(2)_L \times U(1)_Y \times SU(3)_c$  and, 
in addition, it is Hermitian. The latter requirement is imposed on the Lagrangian. In contrast to other terms 
in the Lagrangian  which are naturally Hermitian, the Yukawa interaction ``acquires'' Hermiticity which may not be
necessary. This aspect has been raised in~\cite{Korchin:2016rsf}.

Note that Ref.~\cite{Korchin:2016rsf} also suggested a modification of the SM electroweak interaction to the 
case of a non-Hermitian interaction between the Higgs fields and fermions. The consideration there was restricted to one
generation of the fermions.  It was shown that for positive values of the Yukawa couplings, the fermions get the 
positive mass and the interaction between the Higgs boson and fermions violates $\mathcal{CP}$ symmetry 
without additional Higgs fields. It seems therefore important to investigate further this mechanism 
in the Higgs-boson decays. 

Let us mention that various aspects of non-Hermitian field theories have been studied in 
Refs.~\cite{Alexandre:2015kra,Alexandre:2017fpq,Alexandre:2017foi,Alexandre:2018uol,
Mannheim:2018dur,Alexandre:2018xyy,Millington:2019dfn,Fring:2019hue,Alexandre:2019jdb,Fring:2019xgw,Alexandre:2020wki,Fring:2020bvr,
Alexandre:2020bet,Alexandre:2020tba,Fring:2020xpi,Alexandre:2020gah,Mavromatos:2020hfy,Fring:2020wrj,Mavromatos:2020bbq}.
Influence of non-Hermiticity of the Yukawa interaction on the photon polarization parameters in the $h \to \gamma \gamma$ and $h \to \gamma Z$
decays has been addressed in~\cite{Korchin:2013ifa, Korchin:2013jja}, and on the forward-backward lepton asymmetry in the
$h \to \gamma \ell^+ \ell^-$ ($\ell = e, \mu, \tau$) decays  in~\cite{Korchin:2014kha, Kovalchuk:2017wcf}.

In the present paper, we investigate the decay of the Higgs boson  to a pair of the $\tau$ leptons with their consequent 
decay through the $\tau \to \pi \nu_\tau$ channel, namely  the $h\to\tau^- \tau^+\to\pi^-\nu_{\tau}\,\pi^+{\bar \nu}_\tau$ process.
The case of a non-Hermitian interaction of the Higgs boson with the $\tau$ leptons is considered.

In Sec.~\ref{sec:formalism}, the full angular distribution of the pions is obtained. Then we derive the distribution of the 
angle between the $ \tau^- \to \pi^-\nu_{\tau}$ and $\tau^+ \to \pi^+ {\bar \nu}_{\tau}$ decay planes, and the distribution 
of the polar angle of one of the pions in the helicity frame. The influence of non-Hermiticity of the $h \tau \tau $ interaction on 
the pion distributions is calculated and analyzed. In Sec.~\ref{sec:conclusions} the conclusions are presented.


\section{ Angular distributions of pions }
\label{sec:formalism}

We assume that the interaction of the Higgs boson ($h$) with the $\tau$ leptons
is determined by the Lagrangian which includes scalar ($S$) and pseudoscalar ($PS$) parts
\begin{equation}\label{eq:2001}
{\cal L}_{h\tau\tau}=- \frac{m_\tau}{v}\,h\,{\bar
\psi_\tau}\left(a_\tau+i\,b_\tau \gamma_5\right)\psi_\tau \,,
\end{equation}
where $\psi_\tau$ is the field of the fermion, $v=\left(\sqrt{2}G_{\rm F}\right)^{-1/2}\approx 246$ GeV is the
vacuum expectation value of the Higgs field,  $G_F =1.1663787(6) \times 10^{-5}$  GeV$^{-2}$ is the Fermi constant~\cite{Zyla:2020zbs},
$m_\tau$ is the fermion mass and $a_\tau$, $b_\tau$ are complex parameters ($a_\tau=1$ and $b_\tau=0$ correspond to the SM).
Eq.~(\ref{eq:2001}) can be considered as a phenomenological parametrization of effects of new
physics~\cite{Grzadkowski:1995rx,Bernreuther:1997af,Korchin:2016rsf}. For the real-valued parameters $a_\tau$, $b_\tau$, the interaction
(\ref{eq:2001}) is Hermitian, however, we are interested in the case of non-Hermitian interaction with
complex-valued parameters $a_\tau$, $b_\tau$. At the same time, the Higgs interaction with the $W^\pm$ and $Z$ bosons is chosen
Hermitian as in the SM. 

Let us consider the decay of $h$ to a pair of $\tau$ leptons with their consequent decay to the  
$\tau^- \to \pi^- \nu_\tau$ and $\tau^+ \to \pi^+ {\bar\nu}_\tau$ channels.
The differential decay rate  of $h\to\tau^- \tau^+\to\pi^-\nu_{\tau}\,\pi^+{\bar \nu}_\tau$  in the Higgs boson
rest frame can be written as
\begin{eqnarray}
&& \frac{d^3\Gamma(h\to\tau^-\tau^+\to\pi^- \nu_\tau \pi^+{\bar
\nu}_\tau)}{d\cos\theta_-d\cos\theta_+ d\chi}=\Gamma(h\to \tau^-
\tau^+) \nonumber \\ 
 && \qquad \qquad  \times \Bigl( {\rm BR} (\tau \to \pi \nu_\tau)
\Bigr)^2\frac{d^3W}{d\cos\theta_-d\cos\theta_+ d\chi}.
\label{eq:2002}
\end{eqnarray}
\begin{figure}[tbh]
\begin{center}
\includegraphics[width=0.7\textwidth]{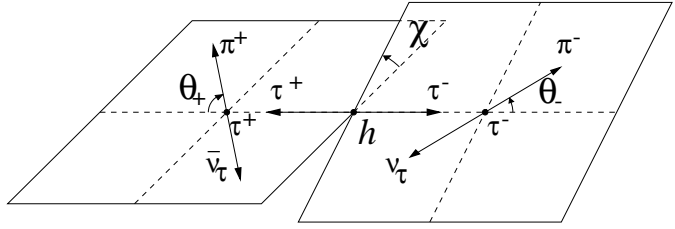}
\end{center}
\caption{Definition of helicity angles $\theta_-$, $\theta_+$, and
$\chi$ in the sequential decay $h\to\tau^- \tau^+\to\pi^-\nu_{\tau}\,\pi^+{\bar \nu}_\tau$ .}
\label{fig:angles}
\end{figure}
Here, $\Gamma(h\to \tau^- \tau^+)$ is the Higgs boson decay width which in the leading order
is given by
\begin{equation}
\Gamma (h \to \tau^-\tau^+)\, = \,m_h \, \beta_\tau \frac{
G_F\,m_\tau^2}{4 \sqrt{2} \pi} \,  \bigl( |a_\tau|^2 \beta_\tau^2
\, + \, |b_\tau|^2 \bigr) \,, \label{eq:2003}
\end{equation}
where $m_h$  is the Higgs boson mass, $\beta_\tau = \sqrt{1-4m_\tau^2/m_h^2}$ is the velocity of 
the $\tau$ lepton in the rest frame of the Higgs, ${\rm BR} (\tau \to \pi \nu_\tau)$ is the branching of 
the $\tau$ decay through the $\tau \to \pi \nu_\tau $ channel. Further,  the total angular distribution
for the $h\to\tau^- \tau^+\to\pi^-\nu_{\tau}\,\pi^+{\bar \nu}_\tau$  decay has the form of
\begin{eqnarray}
&& \frac{d^3\,W}{d\cos\theta_- d\cos\theta_+
d\chi} = \frac{1}{8\pi}\Bigl(1-\cos\theta_-\cos\theta_+  \nonumber \\
&& \qquad - \frac{2\,{\rm Im}(a_\tau\,b_\tau^*)}{|a_\tau|^2 \beta_\tau^2 +
|b_\tau|^2}\,\beta_\tau\,(\cos\theta_--\cos\theta_+ ) \nonumber
\\
&& \qquad -\frac{|a_\tau|^2\beta_\tau^2-|b_\tau|^2}{|a_\tau|^2\beta_\tau^2+|b_\tau|^2}
\sin\theta_-\sin\theta_+\cos\chi  \nonumber \\
&& \qquad -\frac{2\,{\rm Re}(a_\tau
\,b_\tau^*)}{|a_\tau|^2\beta_\tau^2+|b_\tau|^2}\beta_\tau\,\sin\theta_-\sin\theta_+\sin\chi\Bigr),
\label{eq:2004}
\end{eqnarray}
where $\theta_-$ ($\theta_+$) is the angle between the direction of the $\pi^-$ ($\pi^+$) meson
motion in the $\tau^-$ ($\tau^+$) lepton rest frame and the direction of the $\tau^-$ ($\tau^+$) lepton motion in the $h$
boson rest frame, and $\chi$ is the angle between the planes of the  $\tau^-\to \pi^- \nu_\tau$ and $\tau^+\to \pi^+{\bar
\nu}_\tau$  decays in the $h$ boson rest frame (see Fig.~\ref{fig:angles}).

It is useful in parameterization of Eq.~(\ref{eq:2004}),  instead of parameters $a_\tau$, $b_\tau$,
to introduce the parameters (angles) $\phi_{\rm CP}$ and $\phi_{\rm H}$ defined as
\begin{equation}
\tan\phi_{\rm
CP}\equiv\frac{|b_\tau|}{|a_\tau|}\,,\label{eq:2005a}
\end{equation}
\begin{equation}
\frac{2\,{\rm Im}(a_\tau\,b_\tau^*)}{|a_\tau|^2 +
|b_\tau|^2}=\sin2\phi_{\rm CP}\sin\phi_{\rm H}\,,\label{eq:2005b}
\end{equation}
\begin{equation}
\frac{2\,{\rm Re}(a_\tau\,b_\tau^*)}{|a_\tau|^2 +
|b_\tau|^2}=\sin2\phi_{\rm CP}\cos\phi_{\rm H}\,. \label{eq:2005c}
\end{equation}
As a result, Eq.~(\ref{eq:2004}) takes the form of
\begin{eqnarray}
&& \frac{d^3\,W}{d\cos\theta_-d\cos\theta_+
d\chi} = \frac{1}{8\pi}\Bigl(1-\cos\theta_-\cos\theta_+ \nonumber \\
&& \quad - \frac{\beta_\tau\sin2\phi_{\rm CP}\sin\phi_{\rm
H}}{\beta_\tau^2\cos^2\phi_{\rm CP}+ \sin^2\phi_{\rm
CP}}\bigl(\cos\theta_--\cos\theta_+\bigr)\nonumber \\
&& \quad - \Bigl(\frac{\beta_\tau^2\cos^2\phi_{\rm CP}
- \sin^2\phi_{\rm CP}}{\beta_\tau^2\cos^2\phi_{\rm CP}+
\sin^2\phi_{\rm CP}} \cos\chi  \nonumber \\ 
&& \quad +\frac{\beta_\tau\sin2\phi_{\rm
CP}\cos\phi_{\rm H}}{\beta_\tau^2\cos^2\phi_{\rm CP}+
\sin^2\phi_{\rm
CP}}\,\sin\chi\Bigr)\sin\theta_-\sin\theta_+\Bigr).
\label{eq:2006}
\end{eqnarray}
As the $\tau$ leptons produced in the decay of Higgs boson of the mass of 125 GeV are ultrarelativistic, 
one has $\beta_\tau\approx 0.9996$, and taking the limit $\beta_\tau\to 1$, we obtain
\begin{eqnarray}
&& \frac{d^3\,W}{d\cos\theta_-d\cos\theta_+
d\chi}=\frac{1}{8\pi}\Bigl(1-\cos\theta_-\cos\theta_+  \nonumber \\
&& \qquad - \sin2\phi_{\rm CP}\sin\phi_{\rm
H}\bigl(\cos\theta_--\cos\theta_+\bigr)\nonumber \\
&& \qquad - \Bigl(\cos2\phi_{\rm CP}\cos\chi+\sin2\phi_{\rm
CP}\cos\phi_{\rm H}\sin\chi\Bigr) \nonumber \\
&& \qquad \times \sin\theta_-\sin\theta_+\Bigr).
\label{eq:2007}
\end{eqnarray}

For Hermitian $h \tau \tau$ interaction, $\phi_{\rm H}=0$ or $\phi_{\rm H}=\pi$,  Eq.~(\ref{eq:2007}) becomes
\begin{eqnarray}
 &&\frac{d^3\,W}{d\cos\theta_-d\cos\theta_+
d\chi}=\frac{1}{8\pi}\Bigl(1-\cos\theta_-\cos\theta_+\nonumber \\
&& \qquad \qquad \quad - \cos\bigl(\chi\pm2\phi_{\rm
CP}\bigr)\sin\theta_-\sin\theta_+\Bigr). \label{eq:2008}
\end{eqnarray}
Therefore, one of the observables with a maximal sensitivity to the correlations of the $\tau $ spins is the azimuthal angular
correlation in the Higgs rest frame, which has a simple form~\cite{Hagiwara:2016zqz}
\begin{equation}
\frac{dW}{d\chi}=\frac{1}{2\pi}\Bigl(1-\frac{\pi^2}{16}\cos\bigl(\chi\pm2\phi_{\rm
CP}\bigr)\Bigr). \label{eq:2009}
\end{equation}

For a non-Hermitian $h \tau \tau $ interaction (\ref{eq:2001}), this angular correlation
takes a different form of
\begin{eqnarray}
\frac{dW}{d\chi}&=&\frac{1}{2\pi}\Bigl(1-\frac{\pi^2}{16}\bigl(\cos2\phi_{\rm
CP}\cos\chi  \nonumber \\
&& +\sin2\phi_{\rm CP}\cos\phi_{\rm
H}\sin\chi\bigr)\Bigr). \label{eq:2010}
\end{eqnarray}
One has to note that if it will not be possible to distinguish in experiments the events with the azimuthal angle 
$\chi$ from the events with the angle $2\pi-\chi$, then the resulting distribution of $\chi$ takes the form of
\begin{equation}
\frac{dW}{d\chi}=\frac{1}{\pi}\Bigl(1-\frac{\pi^2}{16}\cos2\phi_{\rm
CP}\cos\chi\Bigr),  \; \quad   0\leq\chi\leq\pi . \label{eq:2011}
\end{equation}

\begin{figure}[tbh]
\begin{center}
\includegraphics[width=0.6\textwidth]{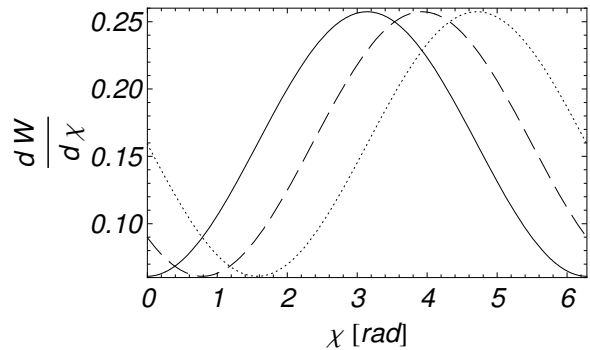}
\end{center}
\caption{Distribution of the azimuthal angle $\chi$  ($0\leq\chi\leq 2\pi$) for Hermitian interaction with $\phi_{\rm H}=0$. 
Solid line corresponds to the SM, dashed line -- $\phi_{\rm CP}=\frac{\pi}{8}$,  dotted line --  $\phi_{\rm CP}=\frac{\pi}{4}$.}
\label{fig:chiCP}
\end{figure}
\begin{figure}[tbh]
\begin{center}
\includegraphics[width=0.6\textwidth]{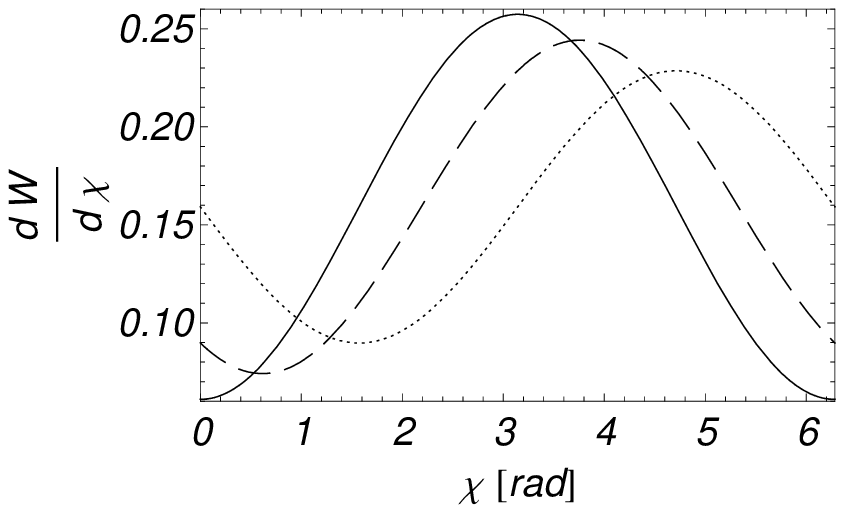}
\end{center}
\caption{Distribution of the azimuthal angle $\chi$.  Solid line -- SM, 
dashed and dotted lines correspond to non-Hermitian interaction with $\phi_{\rm H}=\frac{\pi}{4}$: 
dashed line -- $\phi_{\rm CP}=\frac{\pi}{8}$,  dotted line -- $\phi_{\rm CP}=\frac{\pi}{4}$.}
\label{fig:chinonH1}
\end{figure}
\begin{figure}[tbh]
\begin{center}
\includegraphics[width=0.6\textwidth]{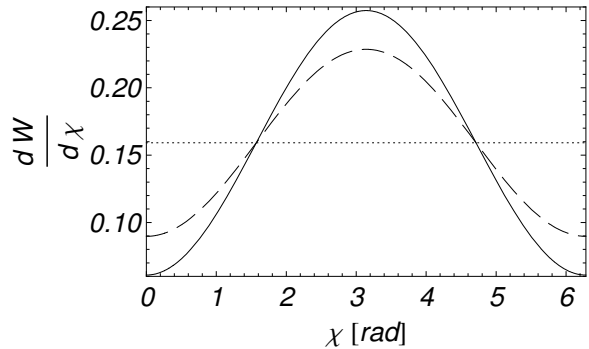}
\end{center}
\caption{Distribution of the azimuthal angle $\chi$.  Solid line -- SM, 
dashed and dotted lines correspond to non-Hermitian interaction with $\phi_{\rm H}=\frac{\pi}{2}$: 
dashed line -- $\phi_{\rm CP}=\frac{\pi}{8}$, dotted line -- $\phi_{\rm CP}=\frac{\pi}{4}$.}
\label{fig:chinonH2}
\end{figure}

In general, a possibility of measurement of the distribution (\ref{eq:2009}) in experiments at the LHC or the ILC 
has been discussed by many authors with the aim of searching for violation of $\mathcal{CP}$ symmetry in the 
$h\to\tau^-\tau^+$ decay (see, for example,~\cite{Hagiwara:2016zqz,Jeans:2018anq}). As for the influence of a 
non-Hermitian interaction (\ref{eq:2001}) on the form of the distribution of the observable $\chi$, this aspect has not been discussed. 

In Figs.~\ref{fig:chiCP},  \ref{fig:chinonH1}, and \ref{fig:chinonH2}, we show the angular distribution (\ref{eq:2010})
for Hermitian and non-Hermitian interactions for a few values of the $\mathcal{CP}$-violation 
parameter $\phi_{\rm CP}$  and Hermiticity-violation parameter $\phi_{\rm H}$.

It is seen from Fig.~\ref{fig:chiCP} that for Hermitian interaction, one can measure violation of 
$\mathcal{CP}$ symmetry via the phase shift in the distribution of $\chi$  (\ref{eq:2009}), if there is no background.
If the interaction is non-Hermitian, then the azimuthal correlation (\ref{eq:2010}) substantially differs
from the SM case, as it is seen in Figs.~\ref{fig:chinonH1} and \ref{fig:chinonH2}, and the corresponding differences strongly depend 
on the parameter $\phi_{\rm H}$.  A model for a non-Hermitian $h \tau \tau$ interaction in which 
$\phi_{\rm H}=\frac{\pi}{2}$ has been discussed in~\cite{Korchin:2016rsf}.

It would also be interesting to measure the following asymmetries:
\begin{eqnarray}
A_1 &\equiv& \Bigl(\int_0^{\pi/2}d\chi-\int_{\pi/2}^{3\pi/2}d\chi+\int_{3\pi/2}^{2\pi}d\chi\Bigr)\frac{dW}{d\chi} 
 = -\frac{\pi}{8}\cos2\phi_{\rm CP}, \label{eq:2012} \\
A_2 &\equiv& \Bigl(\int_0^{\pi}d\chi-\int_{\pi}^{2\pi}d\chi\Bigr)\frac{dW}{d\chi} 
 = -\frac{\pi}{8}\sin2\phi_{\rm CP}\cos\phi_{\rm H}. \label{eq:2013}
\end{eqnarray}
If the values of these asymmetries turn out to be different from the SM prediction, then
this will be a clear signal of physics beyond the SM.

Finally, we briefly discuss another observable which is sensitive to non-Hermiticity
of the Yukawa interaction.  This is the polar-angle correlation
\begin{equation}
\frac{dW}{d\cos\theta_\pm}=\frac{1}{2}\Bigl(1\pm\sin2\phi_{\rm
CP}\sin\phi_{\rm H}\cos\theta_\pm\Bigr).
\label{eq:2014}
\end{equation}
It follows from Eq.~(\ref{eq:2014}) that for Hermitian interaction ($\phi_{\rm H} =0$ or $\phi_{\rm H}= \pi$), 
the distribution of the observable $\cos\theta_\pm$  is uniform. Therefore, any
deviation of a measured distribution from 1/2  will point to a non-Hermiticity of the Yukawa interaction and 
violation of the $\mathcal{CP}$ symmetry in the Higgs boson decay to a pair of $\tau$ leptons.


\section{Conclusions}
\label{sec:conclusions}

In this work, we analyzed the differential rate of the decay of the Higgs boson
to a pair of $\tau $ leptons with their subsequent decay into the 
$\tau^-\to \pi^- \nu_\tau$ and $\tau^+\to \pi^+{\bar \nu}_\tau$ channels.
The Yukawa interaction between the Higgs boson and $\tau$ leptons is assumed to include both the scalar ($S$)
and pseudoscalar ($PS$) couplings.
The total angular distribution of the pions in the $h \to \tau^- \tau^+ \to \pi^- \nu_{\tau}\,\pi^+ {\bar \nu}_\tau$ process  
is considered, as well as distribution of the angle $\chi$ between
the planes spanned by the $\tau^-\to \pi^- \nu_\tau$ and $\tau^+\to \pi^+{\bar \nu}_\tau$ decays,
and distribution of the polar angle $\theta_{\pm}$ of the $\pi^{\pm}$.

For real values of the $S$ and $PS$ couplings, this decay is known to be a source of information on $\mathcal{CP}$ violation in
the $h \tau \tau $ interaction. In the present work, we concentrate on a non-Hermitian Yukawa interaction. 
It is shown that the distributions of the charged pions strongly depend on a parameter of non-Hermiticity 
of the $h \tau \tau$ interaction. Asymmetries sensitive to parameters of $\mathcal{CP}$ violation and 
non-Hermiticity are proposed.

In summary, measurement of the  $h\to\tau^- \tau^+\to\pi^-\nu_{\tau}\,\pi^+{\bar \nu}_\tau$ decay
allows one to test predictions of the SM, and can be a source of information on $\mathcal{CP}$ violation 
in the Yukawa interaction and on such a fundamental property as Hermiticity of this interaction.


\section*{Acknowledgments}

This work was partially conducted in the scope of the IDEATE International Associated Laboratory (LIA).
A.Yu.K. acknowledges partial support by the National Academy of Sciences of Ukraine 
via the programs ``Support for the development of priority areas of scientific research'' (6541230)
and ``Participation in the international projects in high energy and nuclear physics''  
(project No. 0121U111693). 



\end{document}